\documentclass[runningheads]{llncs}
\usepackage{booktabs} 
\usepackage{multirow}
\usepackage{graphicx}
\usepackage{amsmath}
\usepackage{enumitem}
\usepackage[capitalise]{cleveref}
\usepackage{tikz}
\usetikzlibrary{positioning}
\usetikzlibrary{calc}
\usetikzlibrary{shapes.geometric}
\tikzset{database/.style={cylinder,aspect=0.5,draw,rotate=90,path picture={
\draw (path picture bounding box.160) to[out=180,in=180] (path picture bounding
box.20);
\draw (path picture bounding box.200) to[out=180,in=180] (path picture bounding
box.340);
}}}
\begin{document}
\title{Analysing Diffusion Segmentation for Medical Images}
%
%\titlerunning{Abbreviated paper title}
% If the paper title is too long for the running head, you can set
% an abbreviated paper title here
%
%\author{anonymous}
\author{Mathias Öttl\inst{1} \and Siyuan Mei\inst{1} \and Frauke Wilm\inst{1,4} \and Jana Steenpass\inst{2} \and Matthias Rübner\inst{3} \and Arndt Hartmann\inst{2} \and Matthias Beckmann\inst{3} \and Peter Fasching\inst{3} \and Andreas Maier\inst{1} \and Ramona Erber\inst{2} \and Katharina Breininger\inst{4}}
\authorrunning{M.~Öttl et al.}
% First names are abbreviated in the running head.
% If there are more than two authors, 'et al.' is used.
%
%\institute{anonymous}
\institute{Pattern Recognition Lab, Friedrich-Alexander-Universität Erlangen-Nürnberg (FAU), Germany \and Institute of Pathology, University Hospital Erlangen, FAU, Germany \and Department of Gynecology and Obstetrics, University Hospital Erlangen, FAU, Germany \and Department Artificial Intelligence in Biomedical Engineering, FAU, Germany}
\maketitle              % typeset the header of the contribution

\begin{abstract}

Denoising Diffusion Probabilistic models have become increasingly popular due to their ability to offer probabilistic modeling and generate diverse outputs. This versatility inspired their adaptation for image segmentation, where multiple predictions of the model can produce segmentation results that not only achieve high quality but also capture the uncertainty inherent in the model. 
Here, powerful architectures were proposed for improving diffusion segmentation performance.
However, there is a notable lack of analysis and discussions on the differences between diffusion segmentation and image generation, and thorough evaluations are missing that distinguish the improvements these architectures provide for segmentation in general from their benefit for diffusion segmentation specifically.
In this work, we critically analyse and discuss how diffusion segmentation for medical images differs from diffusion image generation, with a particular focus on the training behavior. Furthermore, we conduct an assessment how proposed diffusion segmentation architectures perform when trained directly for segmentation. Lastly, we explore how different medical segmentation tasks influence the diffusion segmentation behavior and the diffusion process could be adapted accordingly.
With these analyses, we aim to provide in-depth insights into the behavior of diffusion segmentation that allow for a better design and evaluation of diffusion segmentation methods in the future.

\end{abstract}

\section{Introduction}

Diffusion models have emerged as powerful tools across various computer vision tasks, including image generation~\cite{diff_beat_gan}, object detection~\cite{diff_object_detect}, and segmentation~\cite{diff_seg}. Their ability to capture complex distributions and generate diverse outputs has made them particularly appealing for applications requiring probabilistic modeling. In recent years, there has been a notable leap towards utilizing diffusion models for segmentation tasks, driven by the motivation to leverage their capabilities in modeling uncertainty. This is especially relevant in the medical domain, where label noise is common and measures of uncertainty are desirable.

However, despite the promise of diffusion models, their application to segmentation tasks poses unique challenges. Unlike natural images, segmentations lack fine-grained details and exhibit different properties that may not be adequately addressed by diffusion schedules designed for image generation.
Efforts to improve diffusion-based segmentation include refining pixel-wise loss functions~\cite{diff_boundary1,diff_boundary2} and developing specialized architectures~\cite{diff_seg,diff_medSeg,diff_medSeg_v2}. 

While specialized architectures claim improved segmentation performance, their diffusion specific advantage has not been thoroughly evaluated. Also, to the best of the authors' knowledge, no prior works analyzed the differences between diffusion image generation and diffusion segmentation. This includes the behavior of different datasets. % and whether the diffusion process needs to be adapted to each dataset. 

In this work, we address three topics. First, we compare three diffusion segmentation architectures in a diffusion segmentation vs.\ feed-forward segmentation setup and visualize the model uncertainty present for different networks and trainings. Second, we discuss the differences of diffusion image generation and diffusion segmentation and how the training behavior might be different. Third, we analyze the forward process of three datasets and suggest how diffusion segmentation methods could be adapted to varying datasets.

%In contrast to a paper proposing a novel methodology, our aim is to provide novel insights into an in vogue approach; due to this, we structure the paper with a focus on the dataset and experiments we use as evidence for our analyzes. 

\section{Background on Diffusion Models}

Diffusion models are generative models that have shown great success in image generation~\cite{diff_beat_gan}. They encompass a forward process, where data is gradually degraded, and a reverse process, where the degradation is reduced. The forward process is described as a Markov Chain with $T$ steps, where a noise estimation model $\epsilon_{\theta}$ is trained to estimate the noise $\epsilon$ that was added onto the input image $x_{0}$. By estimating the noise $\epsilon$ or, alternatively, $x_{0}$, at each timestep $t$ the objective function can be described in two ways as either
\begin{align}
    L_{DM}&={E}_{x, \epsilon \sim \mathcal{N}(0,1), t}\left[\left\|\epsilon-\epsilon_\theta\left(x_t, t\right)\right\|_2^2\right] \enspace \text{or}
    \label{eq:diff_img_gen} \\
%\end{equation}
%
%One variation of the training objective for diffusion models is when training objective is to estimate $x_{0}$ from $x_{t}$ instead of the noise $\epsilon$. In this case the training objective is formulated as
%\begin{equation}
    L_{DM'}&={E}_{x, \epsilon \sim \mathcal{N}(0,1), t}\left[\left\|x_{0}-\epsilon_\theta\left(x_t, t\right)\right\|_2^2\right] \enspace .
    \label{eq:diff_mask_rec}
\end{align}

In addition to learning the data distribution $p(x)$, diffusion models are also capable to model conditional distributions $p(x|y)$ by incorporating the condition $y$ into the training objective as follows:
\begin{equation}
    L_{CDM}={E}_{x, \epsilon \sim \mathcal{N}(0,1), t}\left[\left\|\epsilon-\epsilon_\theta\left(x_t, t, y\right)\right\|_2^2\right] \enspace .
    \label{eq:diff_training}
\end{equation}

For the case of diffusion segmentation, the input $x$ is the segmentation mask and the condition $y$ is the conditioning image to be segmented.

\section{Datasets and Experiments}

%This work will be based on analyzes and discussion of the diffusion process for medical image segmentation and diffusion segmentation architectures. Since these will be based on or involve experiment results, we first introduce the datasets and experiment setups used in the later section of this work.

\subsection{Datasets}

We utilize two publicly available medical datasets, ISIC16~\cite{isic16_ds}, MoNuSeg~\cite{MoNuSeg_1,MoNuSeg_2}, which both have been used in related work for diffusion segmentation, as well as one internal dataset, HER2, for this work.
%Three medical datasets for segmentation are part of this work, with two having been used for diffusion segmentation before. 

\noindent\textbf{ISIC16}~\cite{isic16_ds} Task 3B dataset: This dataset consists of dermoscopic images with the task of lesion segmentation, and was previously used in~\cite{diff_medSeg_v2}. Each image in the dataset contains a lesion, which is typically in the center of the image.

\noindent\textbf{MoNuSeg}~\cite{MoNuSeg_1,MoNuSeg_2} dataset: This dataset consists of H\&E stained images, with labels for multi-organ nuclei segmentation, and was previously used in~\cite{diff_seg}. Each image in the dataset contains multiple nuclei and their segmentations.

\noindent\textbf{HER2} dataset: This is a private dataset comprising of HER2-stained breast cancer tissue biopsy images. The task for this dataset is tumor segmentation, which consists of cell clusters of varying sizes. Tumor tissue might be absent in some images, while covering the majority of an image in others.

\subsection{Experiments}

We cover three architectures that have been proposed for diffusion segmentation:
\begin{itemize}[topsep=0pt]
\item\textbf{EnsemDiff}, introduced in \cite{diff_ensem}, adopts a UNet-like structure, where the conditioning images are concatenated to the noise input.
\item\textbf{SegDiff}, proposed by \cite{diff_seg}, employs its own encoder to process the conditioning images and adds the resulting embeddings at each resolution level to the encoder features.
\item\textbf{MedSegDiff}, proposed by \cite{diff_medSeg}, also utilizes an encoder to process conditioning images but incorporates feature merging at the intermediate resolution levels with feature frequency parsers.
\end{itemize}
\noindent We use these networks in four sets of experiments:

\noindent\textbf{E1) Feed-forward Segmentation}, where training is performed in the classical way with a Dice and cross-entropy loss using the image as input. While recent works reported the diffusion segmentation performance of their proposed architectures, results of the same architectures for feed-forward segmentation are missing. Such comparisons are important to understand the diffusion-specific benefit of an architecture. For this experiment, we also include the SegFormer-B3~\cite{segformer} and nnUnet~\cite{nnUNet} as additional reference architectures.

\noindent\textbf{E2) Diffusion Segmentation}, where the three diffusion segmentation architectures are trained in their intended manner acc. to \cref{eq:diff_training}, predicting $\epsilon$.

\noindent\textbf{E3) Mask Prediction}. For natural images, this degradation usually corresponds to a loss of information, since details are gradually lost. In diffusion segmentation, where a segmentation mask is the target $x$, the loss of information is not continuous, since segmentation mask do not contain fine-grained details that are lost in lower noise levels. We aim to understand how this property impacts the diffusion behavior. Therefore, this experiment encompasses training the three diffusion segmentation networks on recovering the segmentation mask $x_{0}$ from the noisy mask $x_{t}$ independent of the image input, according to \cref{eq:diff_mask_rec}. %Diffusion models are trained to predict the noise $\epsilon$ that degraded an image $x$. 

\noindent\textbf{E4) Image Generation}, represents the classical unconditional image generation according to \cref{eq:diff_img_gen} with the image being $x$.

Settings and training schemes for the diffusion segmentation architectures were adopted from their published works. EnsemDiff provided only one set of settings, for SegDiff we utilized the settings reported for the MoNuSeg experiments, while we took the settings for the ISIC16 dataset for MedSegDiff.
None of the diffusion segmentation works described a training-validation split or any other method for model selection. To use the same amount of training data and to avoid test-set optimization by selecting a metric based on test metrics, we trained the architectures for the reported amount of steps. 
Segmentation results were obtained by ensembling of predictions. For our datasets, we ensembled 25 times for MoNuSeg, 10 times for ISIC16 and 5 times for HER2.

Feed-forward segmentation experiments were performed with an equal combination of Dice and cross-entropy loss, a learning rate of 1e-5 and Adam optimizer. The required input $x_{t}$ of the diffusion segmentation architectures was replaced with random noise of the required shape. To utilize the same amount of training data as for diffusion segmentation, we first trained with an 80-20 train-validation split to determine the number of training epochs and then retrained with the full dataset without validation. 
%We also included nnUNet~\cite{nnUNet} with the default framework settings.
As metrics we utilize Intersection over Union (IoU) and the Expected Calibration Error (ECE)~\cite{ece} with ten bins.

\section{Evaluation}%ing Diffusion Segmentation Architectures}

\subsection{Feed-forward Segmentation Performance vs.\ Diffusion Segmentation}

In this section, we compare results of experiments E1) and E2) with regard to segmentation performance and uncertainty quantification to investigate the impact of the architecture as well as the potential of diffusion models to provide better model calibration.

\begin{table}[tb]
\centering
\resizebox{0.8\textwidth}{!}{%
\begin{tabular}{clccccccccccc}
\toprule
& Method & IoU$\uparrow$ && ECE$\downarrow$ && IoU$\uparrow$ && ECE$\downarrow$ && IoU$\uparrow$ && ECE$\downarrow$ \\
\cmidrule{3-3}\cmidrule{5-5}\cmidrule{7-7}\cmidrule{9-9}\cmidrule{11-11}\cmidrule{13-13}
&& \multicolumn{3}{c}{Feed-forward Segmentation} && \multicolumn{3}{c}{Diffusion Segmentation} && \multicolumn{3}{c}{Difference} \\
\cmidrule{3-5}\cmidrule{7-9}\cmidrule{11-13}
\multicolumn{1}{c}{\multirow{5}{*}{\rotatebox[origin=c]{90}{\textbf{ISIC16}}}} 
& nnUNet & 84.48 && 0.147 &&  &&  &&  &&  \\
& SegFormer & 88.34 && 0.024 &&  &&  &&  &&  \\
& EnsemDiff & 84.92 && 0.019 && 86.59 && 0.007 && +1.67 && -0.012 \\
& SegDiff & 85.86 && 0.020 && 86.39 && 0.017 && +0.53 && -0.003 \\
& MedSegDiff & 83.22 && 0.029 && 83.73 && 0.017 && +0.51 && -0.012 \\
%\cmidrule{3-13}
\cmidrule{3-5}\cmidrule{7-9}\cmidrule{11-13}

\multicolumn{1}{c}{\multirow{5}{*}{\rotatebox[origin=c]{90}{\textbf{MoNuSeg}}}} 
& nnUNet & 71.43 && 0.050 &&  &&  &&  &&  \\
& SegFormer & 62.95 && 0.055 &&  &&  &&  &&  \\
& EnsemDiff & 63.90 && 0.061 && 64.64 && 0.041 && +0.74 && -0.020 \\
& SegDiff & 63.44 && 0.059 && 64.57 && 0.038 && +1.13 && -0.021 \\
& MedSegDiff & 62.51 && 0.055 && 63.05 && 0.085 && +0.54 && +0.030 \\

%\cmidrule{3-13}
\cmidrule{3-5}\cmidrule{7-9}\cmidrule{11-13}

\multicolumn{1}{c}{\multirow{5}{*}{\rotatebox[origin=c]{90}{\textbf{HER2}}}} 
& nnUNet & 80.00 && 0.104 &&  &&  &&  &&  \\
& SegFormer & 80.36 && 0.056 &&  &&  &&  &&  \\
& EnsemDiff & 72.57 && 0.075 && 73.70 && 0.018 && +1.13 && -0.057 \\
& SegDiff & 74.90 && 0.063 && 72.80 && 0.011 && -2.10 && -0.052 \\
& MedSegDiff & 70.04 && 0.086 && 67.10 && 0.017 && -2.94 && -0.069 \\

\bottomrule
\end{tabular}
}
\caption{Segmentation results for all methods and datasets used in this work, including the advantage diffusion segmentation has over feed-forward segmentation under the same architecture.}
\label{table:seg_results}
\end{table}

In \cref{table:seg_results}, we report the segmentation results for all architectures and datasets used in this work. For EnsemDiff, SegDiff and MedSegDiff, we compare their feed-forward segmentation performance without the diffusion process with their diffusion segmentation performance. In most experiments, diffusion segmentation training performed better than feed-forward segmentation training, especially when looking at the ECE. However, for each dataset, either nnUNet or SegFormer performed best.

Our results are roughly in the same range as values reported in related work. For the HER2 dataset we suspect a poor fit of the diffusion methods, caused by the absence of a validation scheme of these methods. Still, for our setup, we see a relative improvement of the diffusion segmentation training in relation to the feed-forward segmentation training for most cases. This shows especially for the ECE, where the diffusion segmentation training performs better for all but one case. 

\begin{figure}[ht]
    \centering
    \resizebox{1.0\textwidth}{!}{%
    \begin{tikzpicture}[ image/.style = {inner sep=0pt, outer sep=0pt}, node distance = 1mm and 1mm]
    \def\imageWidth{3cm}
    \def\maskWidth{2.8cm}
    
    % ISIC
    \node [image] (img0) {\includegraphics[width=\imageWidth]{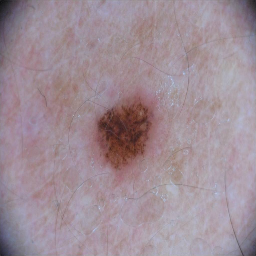}};
    \node [image,right=0.25cm of img0] (mask0) {\includegraphics[width=\imageWidth]{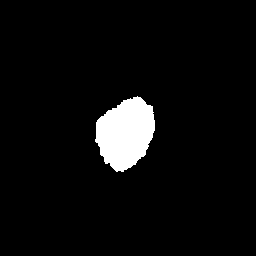}};

    \node[image, right=0.5cm of mask0](diff_mean_0){\includegraphics[width=\imageWidth]{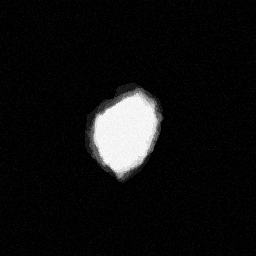}};
    \node[image, right=0.25cm of diff_mean_0](diff_std_0){\includegraphics[width=\imageWidth]{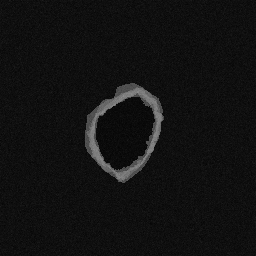}};

    \node[image, right=0.5cm of diff_std_0](seg_mean_0){\includegraphics[width=\imageWidth]{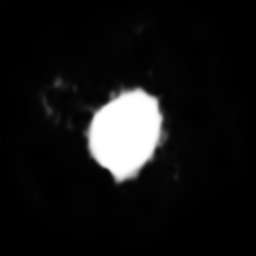}};
    \node[image, right=0.25cm of seg_mean_0](seg_std_0){\includegraphics[width=\imageWidth]{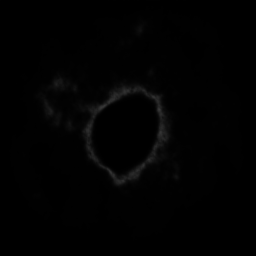}};

    % MoNuSeg
    \node [image,below=0.5cm of img0] (img1) {\includegraphics[width=\imageWidth]{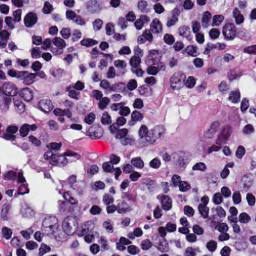}};
    \node [image,right=0.25cm of img1] (mask1) {\includegraphics[width=\imageWidth]{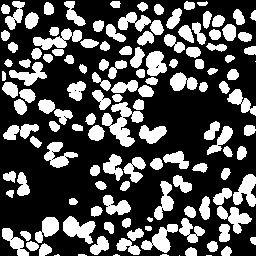}};

    \node[image, right=0.5cm of mask1](diff_mean_1){\includegraphics[width=\imageWidth]{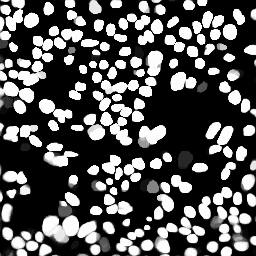}};
    \node[image, right=0.25cm of diff_mean_1](diff_std_1){\includegraphics[width=\imageWidth]{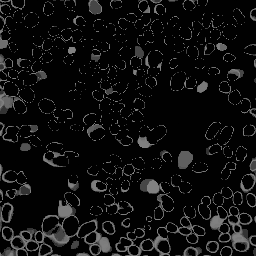}};

    \node[image, right=0.5cm of diff_std_1](seg_mean_1){\includegraphics[width=\imageWidth]{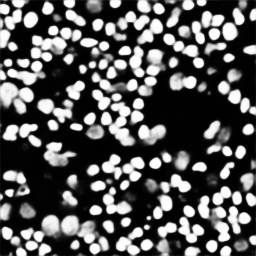}};
    \node[image, right=0.25cm of seg_mean_1](seg_std_1){\includegraphics[width=\imageWidth]{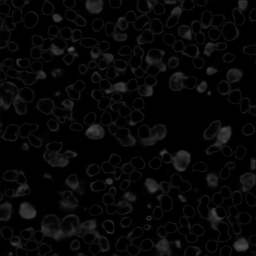}};

    \node[yshift = -2.20cm, inner sep=0] (text1) at ($(img1)$) {\Large{Image}};
    \node[yshift = -2.20cm, inner sep=0] (text2) at ($(mask1)$) {\Large{Ground Truth}};
    \node[yshift = -1.8cm, inner sep=0] (text3) at ($(diff_mean_1)$) {\Large{Mean}};
    \node[yshift = -1.8cm, inner sep=0] (text4) at ($(diff_std_1)$) {\Large{Std}};
    \node[yshift = -1.8cm, inner sep=0] (text5) at ($(seg_mean_1)$) {\Large{Mean}};
    \node[yshift = -1.8cm, inner sep=0] (text6) at ($(seg_std_1)$) {\Large{Std}};

    \node[yshift = -2.4cm, inner sep=0] (text5) at ($(diff_mean_1)!0.5!(diff_std_1)$) {\Large{Diffusion Segmentation}};
    \node[yshift = -2.4cm, inner sep=0] (text6) at ($(seg_mean_1)!0.5!(seg_std_1)$) {\Large{Feed-forward Segmentation}};

    \end{tikzpicture}}
    \caption{Comparision of segmentation results for MoNuSeg with SegDiff (upper row) and ISIC16 with MedSegDiff (lower row), when the networks were trained for diffusion segmentation or for feed-forward segmentation with random noises as $x_{t}$ input.}
    \label{fig:output_variation}
\end{figure}
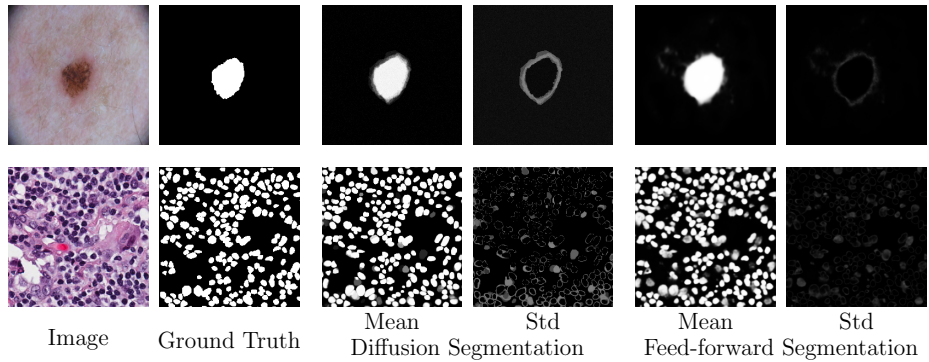

In \cref{fig:output_variation}, we show qualitative segmentation examples for SegDiff (for ISIC16) and MedSegDiff (for MoNuSeg), illustrating the mean and standard deviation of multiple predictions. While the feed-forward segmentation models highlight similar areas as the diffusion models as uncertain, the model is less uncertain overall.
Of note, the uncertainty for the feed-forward segmentation is caused solely by the random noise provided instead of $x_{t}$ during training/inference, which slightly modifies the activations. We expect this to be a ``weak'' form of augmentation/ensembling, as the network can easily learn to ignore this input.

These result hint at the advantage of training the same networks in a diffusion segmentation setup; however, they are -- in our experiments -- not fully competitive with SOTA feed forward networks. Furthermore, more powerful uncertainty / ensemble approaches (including test-time augmentation~\cite{test_time_aug}) may close or negate this gap further.

%\subsection{Uncertainty Expression}

%We have seen that diffusion segmentation shows great success in expressing uncertainty. However, we suspect that networks trained in the classical way can also express a certain amount of uncertainty.

%Classical segmentation can generate varying predictions due to random noise provided for the $x_{t}$ input during training and prediction. Similar variations, but with weaker expression, can be observed.

%We conclude that a similar kind of uncertainty is contained in the network trained for classical segmentation. Providing random noise during inference slightly modifies the activations, creating slightly varying results, although the variations are weaker than those of diffusion segmentation. We suspect this effect could be related to test-time augmentation\cite{test_time_aug} and should be evaluated further to delineate the uncertainty modelling of diffusion segmentation from that of feed-forward segmentation.

%\section{Diffusion Models for Segmentation}
\subsection{Error and Loss Behaviour of Diffusion Segmentation over Different Timesteps}
\label{sec:diff_masks}

\begin{figure*}
    \centering
    \resizebox{1.0\textwidth}{!}{%
    \begin{tikzpicture}[ image/.style = {inner sep=0pt, outer sep=0pt}, node distance = 1mm and 1mm]
    \def\imageWidth{3cm}
    \def\maskWidth{2.8cm}
    
    % Original images
    \node [image] (img) {\includegraphics[width=\imageWidth]{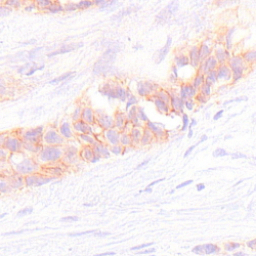}};
    \node [image,below=0.25cm of img] (mask) {\includegraphics[width=\imageWidth]{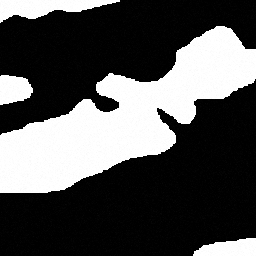}};

    % t = 50
    \node[image, right= 0.5cm of img] (img_50){\includegraphics[width=\imageWidth]{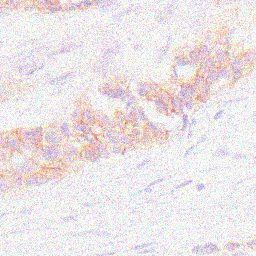}};
    \node[image, right= 0.5cm of mask] (mask_50){\includegraphics[width=\imageWidth]{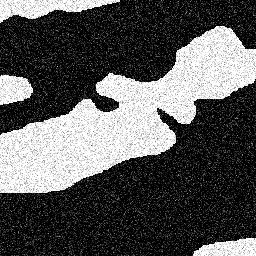}};

    % t = 100
    \node[image, right= 0.25cm of img_50] (img_100){\includegraphics[width=\imageWidth]{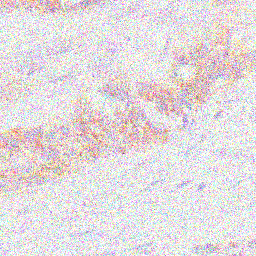}};
    \node[image, right= 0.25cm of mask_50] (mask_100){\includegraphics[width=\imageWidth]{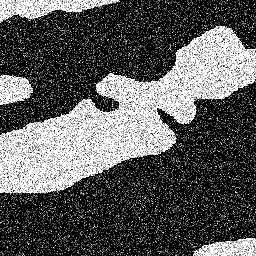}};

    % t = 200
    \node[image, right= 0.25cm of img_100] (img_200){\includegraphics[width=\imageWidth]{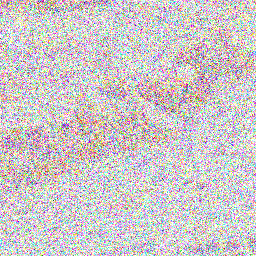}};
    \node[image, right= 0.25cm of mask_100] (mask_200){\includegraphics[width=\imageWidth]{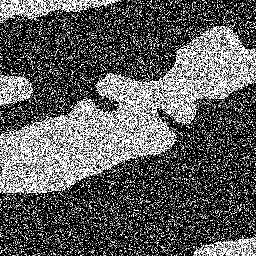}};

    % t = 300
    \node[image, right= 0.25cm of img_200] (img_300){\includegraphics[width=\imageWidth]{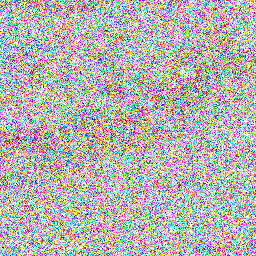}};
    \node[image, right= 0.25cm of mask_200] (mask_300){\includegraphics[width=\imageWidth]{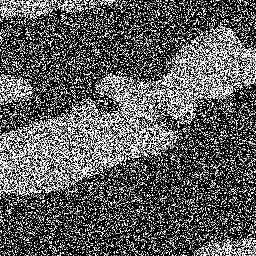}};

    % t = 400
    \node[image, right= 0.25cm of img_300] (img_400){\includegraphics[width=\imageWidth]{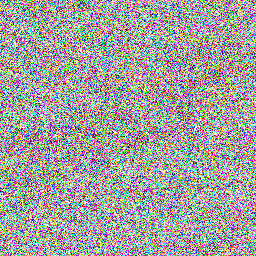}};
    \node[image, right= 0.25cm of mask_300] (mask_400){\includegraphics[width=\imageWidth]{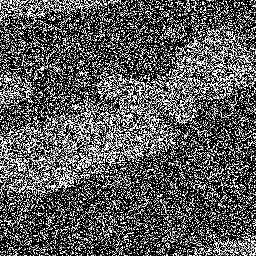}};

    % t = 600
    \node[image, right= 0.25cm of img_400] (img_600){\includegraphics[width=\imageWidth]{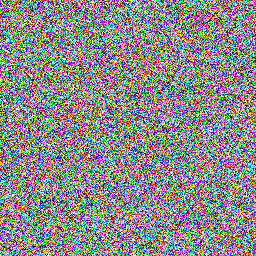}};
    \node[image, right= 0.25cm of mask_400] (mask_600){\includegraphics[width=\imageWidth]{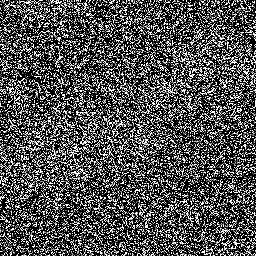}};

    % t = 999
    \node[image, right= 0.25cm of img_600] (img_999){\includegraphics[width=\imageWidth]{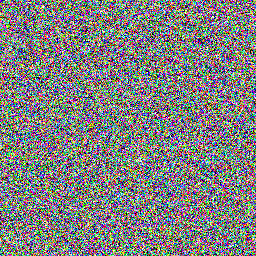}};
    \node[image, right= 0.25cm of mask_600] (mask_999){\includegraphics[width=\imageWidth]{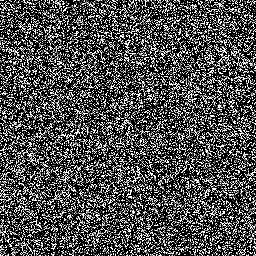}};

    \node[yshift = -2.0cm, inner sep=0] (text1) at ($(mask)$) {\LARGE{t=0}};
    \node[yshift = -2.0cm, inner sep=0] (text2) at ($(mask_50)$) {\LARGE{t=50}};
    \node[yshift = -2.0cm, inner sep=0] (text3) at ($(mask_100)$) {\LARGE{t=100}};
    \node[yshift = -2.0cm, inner sep=0] (text4) at ($(mask_200)$) {\LARGE{t=200}};
    \node[yshift = -2.0cm, inner sep=0] (text5) at ($(mask_300)$) {\LARGE{t=300}};
    \node[yshift = -2.0cm, inner sep=0] (text6) at ($(mask_400)$) {\LARGE{t=400}};
    \node[yshift = -2.0cm, inner sep=0] (text7) at ($(mask_600)$) {\LARGE{t=600}};
    \node[yshift = -2.0cm, inner sep=0] (text8) at ($(mask_999)$) {\LARGE{t=999}};
   
    \end{tikzpicture}}
    \caption{Forward process for images (upper row) and segmentation mask (lower row), on the example of the HER2 dataset, with 1000 diffusion steps and linear beta schedule.}
    \label{fig:degradation}
\end{figure*}
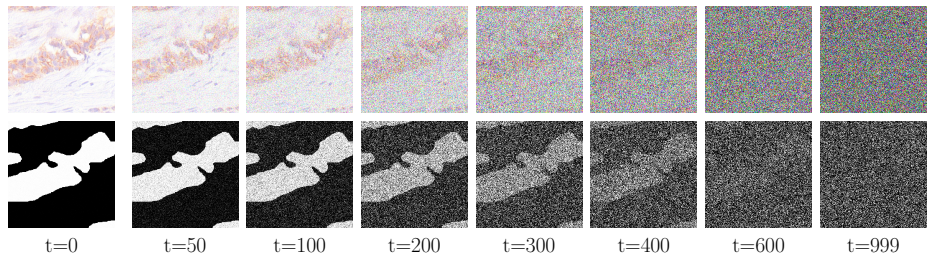

In this section, we investigate how a diffusion segmentation model behaves at different timesteps of the diffusion process. For this, we first compare the normal diffusion segmentation  behavior in E2) with ``unconditioned'' diffusion segmentation from E3 (without an image) that can be understood as mask completion or shape generation. Subsequently, we investigate differences between diffusion segmentation and image generation in terms of the loss function.

\begin{figure}[h]
\centering
 \includegraphics[width=0.32\textwidth]{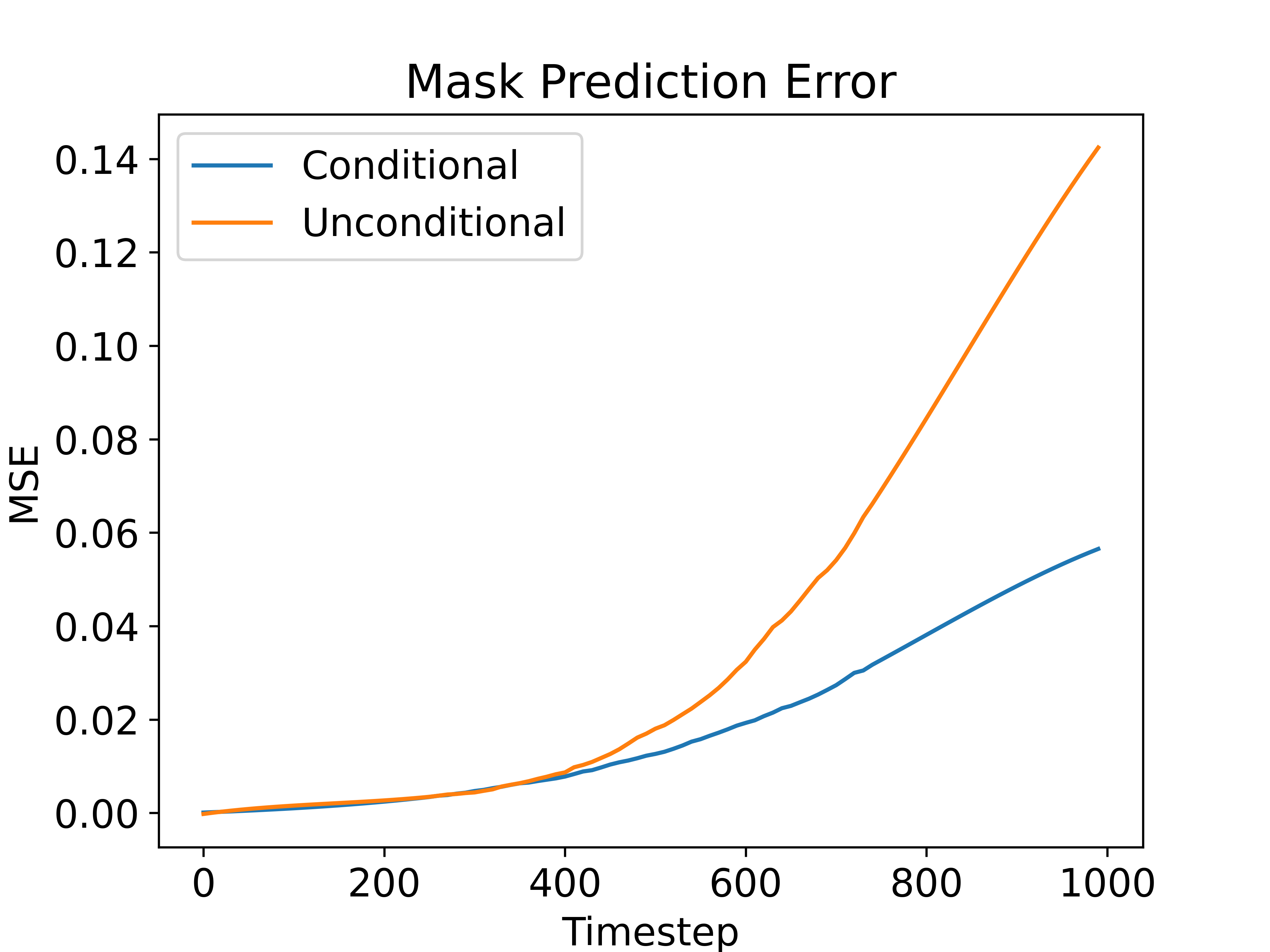}
 \includegraphics[width=0.32\textwidth]{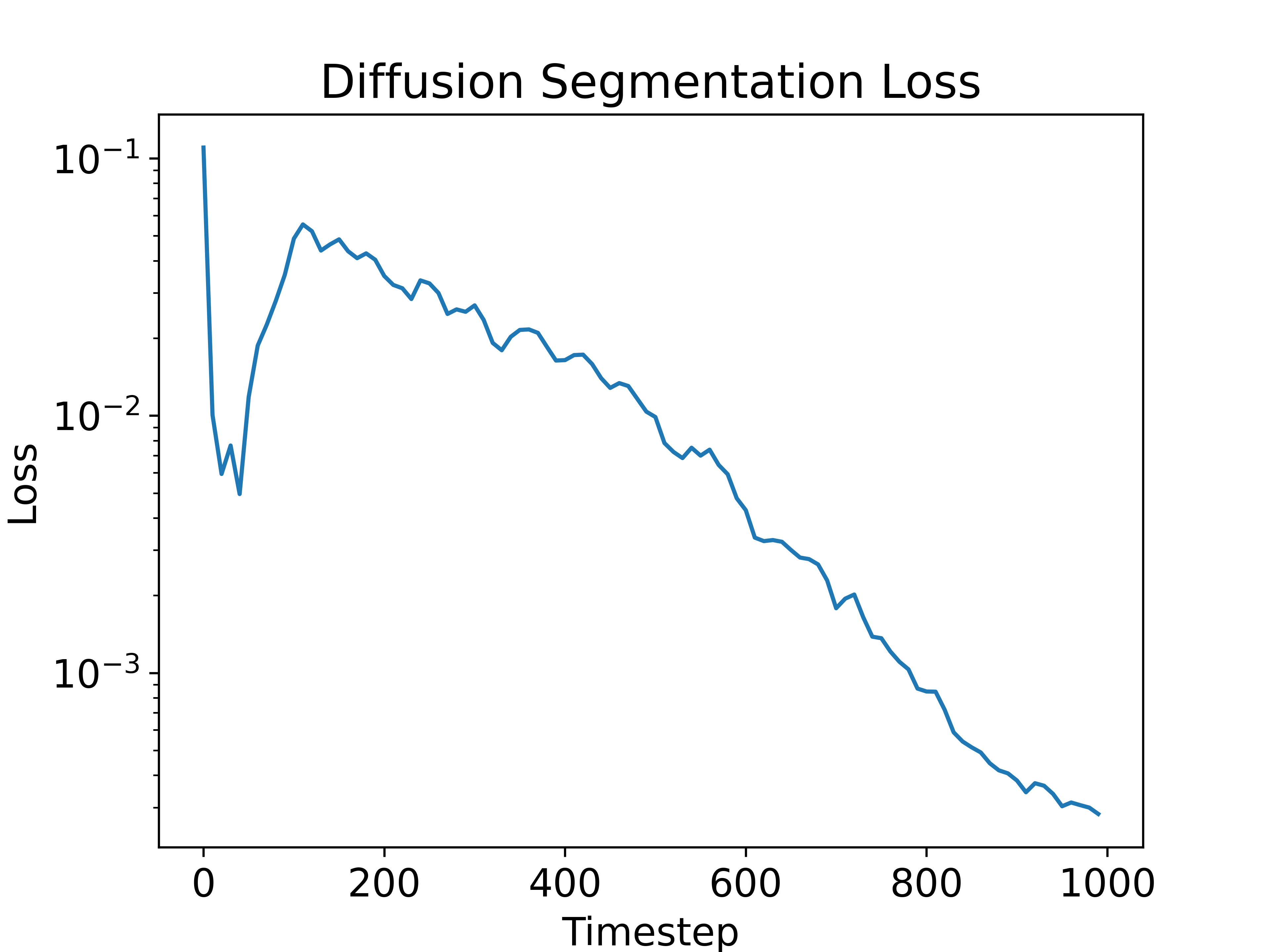}
 \includegraphics[width=0.32\textwidth]{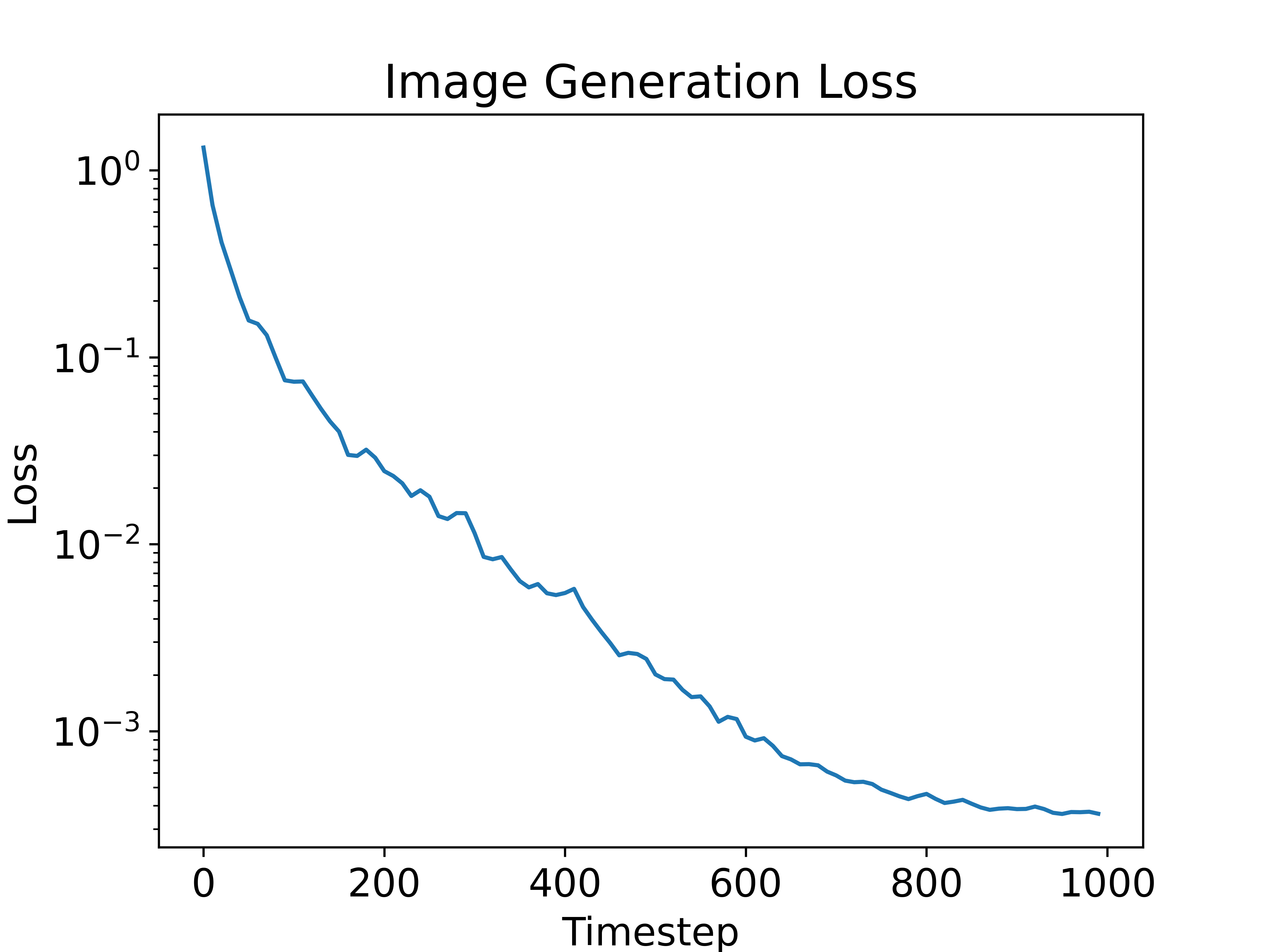}
 \caption{Smoothed curves of the mask prediction error (left), loss for the task of diffusion segmentation (middle) and image generation (right) for EnsemDiff on HER2. Models were picked after 75\% of the total number of training steps.}
 \label{img:training_losses}
\end{figure}

In \cref{img:training_losses} (left), we illustrate the behavior of the mask prediction error on an example from the HER2 dataset. For the first 400 timesteps, we see almost no difference between the conditioned and the unconditioned diffusion segmentation, meaning the model draws no benefit from (and potentially does not pay attention to) the actual image to be segmented for the prediction of the mask for these timesteps. In general, the behavior indicates that the mask $x_{0}$ can be recovered without large errors from $x_{t}$ until this point by simple, content independent, denoising of the mask. 
\cref{fig:degradation} illustrates this effect of increasing noise levels and a potential difference to image denoising on an image (upper row) and a segmentation mask (lower row). While different levels of detail get continuously lost for the image with increasing timesteps, the segmentation mask can be (visually) identified without loss of information until higher timesteps.

%We aim to quantify this effect with the illustration of the mask prediction error of the HER2 dataset in \cref{img:training_losses} (left). For this example, we see a low error until timestep 500, indicating that the mask $x_{0}$ can be recovered without large errors from $x_{t}$ until this point. 
%This could implicate, that during diffusion segmentation training, at lower timesteps $t$, the network can identify the noise $\epsilon$ solely from $x_{t}$ without requiring information from the condition image $y$.

%\subsection{Behaviour of Training Loss}

Following the analysis of the diffusion process on segmentation masks, we now consider the implications on the diffusion training process. 
The middle and right parts of \Cref{img:training_losses} illustrate the training losses for the diffusion segmentation training (E2) and image generation training (E4), respectively. Note that in both cases, the network is trained to predict the noise $\epsilon$. Both loss curves were created with the EnsemDiff architecture using the HER2 data and taken after 75\% of the training steps, where we expect the models to be reasonable fit, while not running into overfitting. 
As expected, the loss over the timesteps for image generation decreases monotonously, as information is continuously lost, but the noise $\epsilon$ is easier to identify. At higher noise levels, the network can forward $x_{t}$ almost unchanged as noise prediction $\epsilon$ and achieve a low loss value as the underlying image $x_{0}$ has decreasing impact. 
For diffusion segmentation, we see a very interesting, nonmonotonous behavior, with a sharp dip within the first 50 timesteps and sharp increase in the next 100 timesteps. For early timesteps, it may be very easy for the network to predict the noise as learning a simple high frequency filter could be sufficient for a decent prediction due to large constant areas in the mask input. As soon as the noise levels prevent this simple estimation, the error sharply increases back to an almost linear behavior. %While the total loss at early timepoints is lower for segmentation compared to generation, this advantage seems to disappear, pointing toward higher difficulties of estimating the noise from the noised input for segmentation compared to image generation.

%In diffusion segmentation, we expect lower loss values for the lower timesteps, where the mask is easy to identify, followed by the continuous decrease at higher timesteps.

%We see the expected behaviour of a monotonously decreasing loss for the image generation. Diffusion segmentation shows a peak of loss at very low timesteps, caused by the difficulty to predict very low noise amplitudes. Afterwards, we see a dip, followed by a flatter loss area, which is the area where the mask is still relatively easy to recover. At higher noise levels we see the same behaviour as for the image generation task.

Based on these observations we argue that the diffusion loss structure for image generation might be ill fit for the task of diffusion segmentation. In \cref{img:training_losses} we showed that the mask $x_{0}$ can be recovered from its noisy version $x_{t}$ without larger error until timestep 400 for this dataset. However, the loss structure indicates that most attention is put towards the lower half of the timesteps, and not on the upper half of the timesteps, where the information from the noisy mask $x_{t}$ starts to degrade and where we expect the probabilistic modelling to happen.
While this does not prevent diffusion segmentation from learning to segment the image, this might cause longer than necessary training times and a suboptimal probabilistic modelling.
Potential improvements for this circumstance could be noise or weight schedules that put more emphasis on the noise levels where the information from the noisy mask $x_{t}$ is starting to degrade. 

\subsection{Dataset Fingerprints}

\begin{figure}[ht]
\centering
 \begin{minipage}[b]{0.65\linewidth}
  \includegraphics[width=\textwidth,trim={0 4px 0 10px},clip]{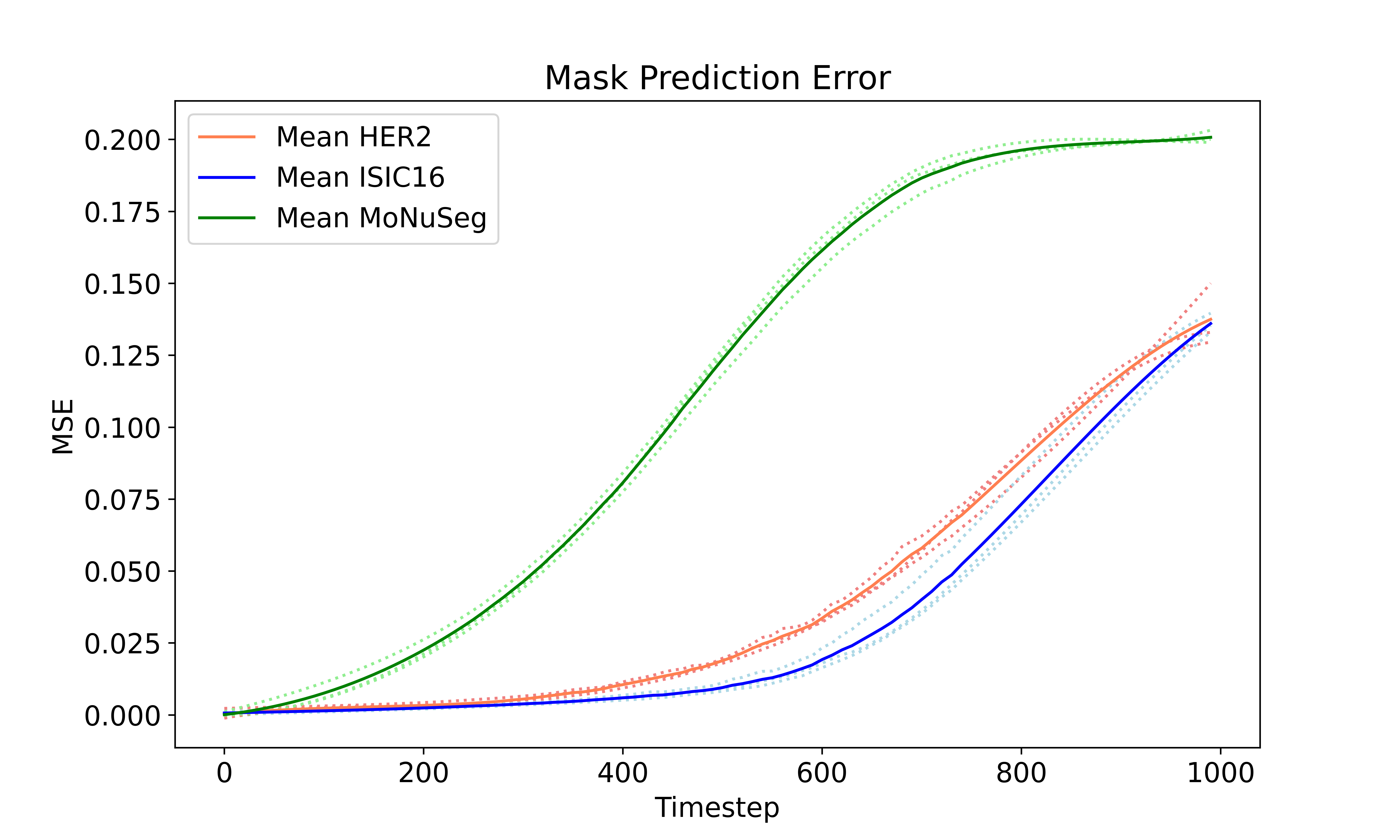}
 \end{minipage}
 \caption{Smoothed mask prediction error for the three datasets for each architecture (dottet) and averaged (solid).}
 \label{img:all_mask_mse}
\end{figure}

\Cref{sec:diff_masks} discussed the diffusion behavior of segmentation mask on the example of the HER2 dataset. However, we suspect that the behavior of diffusion segmentation is depended on a fingerprint of the segmentation task, as the loss of information might not happen similar for each kind of segmentation mask.

In \cref{img:all_mask_mse} we show the smoothed mask prediction error for all three datasets and all three architectures (dotted), as well as the average over the architectures (solid). Different behaviors can be observed for each dataset. While the MSE for the MoNuSeg dataset increases much faster than for the HER2 dataset, it stays low until higher timesteps for the ISIC16 dataset.

We argue this behavior to be caused by the properties of the segmentation task. The nuclei segmentation for MoNuSeg encompasses small objects, which get unrecognizable faster with increasing noise levels. In contrast, the ISIC16 dataset contains one large segmentation mask per image, which is usually located in the center of the image, allowing for an easier recognition even at higher noise levels with little benefit of iterative denoising. %We argue this behaviour to be remotely related to a optical resolution problem. 

We conclude that the segmentation fingerprint might have an impact on the diffusion behavior and should be considered when designing diffusion segmentation methods, especially in the case of medical images, which encompass a large variety of segmentation types. First, not all information appears to be lost at 1000 timesteps for HER2 and ISIC16 since no convergence of the MSE is observed, indicating that higher final noise levels might be required for these datasets. Second, all datasets have timestep areas in which all or no information is recovered from the noisy mask. We propose to put less emphasis on these areas, as no probabilistic modelling should be learned in these areas. A potential improvement could be a weighting schedule with the first deviate of the mask prediction error, which could promote faster training convergence and better modelling of uncertainty.

\section{Conclusion}

In this work, we analyzed and discussed multiple aspects of diffusion segmentation for medical images. We evaluated the advantages of diffusion segmentation training vs.\ feed-forward segmentation training for the same architectures and showed how networks trained for feed-forward segmentation also contain some uncertainty. We provided insights in the different behaviors of diffusion segmentation compared to image generation and how the loss structure might be a poor fit for diffusion segmentation. Lastly, we analyzed the behavior of different medical datasets in a diffusion setting, and discussed how the diffusion process could be adapted to that. 
With this work, we aimed to extend the understanding of diffusion segmentation and propose ideas to potentially adapt diffusion segmentation to diverse medical datasets.

\section{Acknowledgement}
This project is supported by the Bavarian State Ministry of Health and Care, project grants No. PBN-MGP-2010-0004-DigiOnko and PBN-MGP-2008-0003-DigiOnko. We also gratefully acknowledge the support from the Interdisciplinary Center for Clinical Research (IZKF, Clinician Scientist Program) of the Medical Faculty FAU Erlangen-Nürnberg. The authors gratefully acknowledge the scientific support and HPC resources provided by the Erlangen National High Performance Computing Center (NHR@FAU) of the Friedrich-Alexander-Universität Erlangen-Nürnberg (FAU) under the NHR project b160dc. NHR funding is provided by federal and Bavarian state authorities. NHR@FAU hardware is partially funded by the German Research Foundation (DFG) – 440719683. K.B., F.W. and M.Ö. acknowledge support by the German Research Foundation (DFG) project 460333672 CRC1540 EBM. K.B. further acknowledges support by d.hip campus - Bavarian aim in form of a faculty endowment.

%
% ---- Bibliography ----
%
% BibTeX users should specify bibliography style 'splncs04'.
% References will then be sorted and formatted in the correct style.
%
\bibliographystyle{splncs04}
\bibliography{refs}
\end{document}